\documentclass[onecolumn,superscriptaddress]{revtex4-1}
\pdfoutput=1

\renewcommand{\cite}{\citet}

\usepackage{graphicx}
\usepackage{grffile}  

\usepackage[utf8]{inputenc}
\usepackage{mathtools}

\begin{document}

\title{On the reflection of Alfvén waves and its implication for Earth's core modeling}

\author{N. Schaeffer}
\affiliation{ISTerre, CNRS, University Joseph Fourier, BP 53, 38041 Grenoble Cedex 9, France.}
\author{D. Jault}
\affiliation{ISTerre, CNRS, University Joseph Fourier, BP 53, 38041 Grenoble Cedex 9, France.}
\affiliation{Earth and Planetary Magnetism Group, Institut für Geophysik, Sonnegstrasse 5, ETH Zürich, CH-8092, Switzerland}
\author{P. Cardin}
\affiliation{ISTerre, CNRS, University Joseph Fourier, BP 53, 38041 Grenoble Cedex 9, France.}
\author{M. Drouard}
\affiliation{ISTerre, CNRS, University Joseph Fourier, BP 53, 38041 Grenoble Cedex 9, France.}

\date{\today}

\begin{abstract}
Alfvén waves propagate in electrically conducting fluids in the presence of a magnetic field.
Their reflection properties depend on the ratio between the kinematic viscosity and the magnetic diffusivity of the fluid, also known as the magnetic Prandtl number $Pm$.
In the special case $Pm=1$, there is no reflection on an insulating, no-slip boundary, and the incoming wave energy is entirely dissipated in the boundary layer.

We investigate the consequences of this remarkable behaviour for the numerical modeling of torsional Alfvén waves (also known as torsional oscillations), which represent a special class of Alfvén waves, in rapidly rotating spherical shells. They consist of geostrophic motions and are thought to exist in the fluid cores of planets with internal magnetic field. 
In the geophysical limit $Pm \ll 1$, these waves are reflected at the core equator, but they are entirely absorbed for $Pm=1$. Our numerical calculations show that the reflection coefficient at the equator  of these waves remains below $0.2$ for $Pm \ge 0.3$, which is the range of values for which geodynamo numerical models operate. As a result, geodynamo models with no-slip boundary conditions cannot exhibit torsional oscillation normal modes.
\end{abstract}

\maketitle

\section{Introduction}

Hannes Alfvén first showed the theoretical existence, in an inviscid fluid of infinite electrical conductivity, of hydromagnetic waves that couple fluid motion and magnetic field \citep{Alfven1942}.
The propagation of torsional Alfvén waves in the Earth's fluid core was thereafter predicted by \cite{braginski70}.
Such waves arise in rapidly rotating spheres or spherical shells in the presence of a magnetic field.
In torsional Alfv\'en waves, the motions are geostrophic and consist in the rotation
$\omega_g(s)$ of nested cylinders centered on the rotation axis. They thus depend only on the distance $s$ to the rotation axis.
The period of  the fundamental modes of torsional Alv\'en waves in the Earth's fluid core was first estimated to be about 60 years. This timescale was inferred from the analysis of the decadal length of day changes since the first half of the 19th century \citep{jordi1994fluctuations} and of the geomagnetic secular variation after 1900 \citep{braginsky1984short}.
With hindsight, these time series were not long enough to show convincingly variations with 60 years periodicity.
Torsional waves with much shorter periods have now been extracted from time series of core surface flows for the time interval 1955-1985 \citep{Gillet10}.
If this discovery is confirmed, the  period of the fundamental modes is of the order of 6 years and, as such, is much shorter than initially calculated.
 
Several authors have searched for torsional Alfv\'en waves in geodynamo simulations.
Using stress-free boundary conditions, \cite{dumberry03} and \cite{busse_simitev05} illustrated some parts of the torsional wave mechanism. \cite{dumberry03} found that the whole length of the geostrophic cylinders accelerates azimuthally as if they were rigid. The inertial forces, in their simulation, are however so influential that they dominate the Lorentz forces.
Torsional Alfv\'en waves have finally been detected in a set of numerical simulations of the geodynamo with no-slip boundary conditions, for $0.5 \le Pm\le 10$, by \cite{wicht10} (the magnetic Prandtl number $Pm$ is the ratio of kinematic viscosity over magnetic diffusivity).
In both the geophysical ($Pm\sim 10^{-5}$) and the numerical studies, there seems to be no reflection of the torsional Alfv\'en waves upon their arrival at the equator.
However, experimental studies in liquid metals have shown resonance effects on Alfvén normal modes \citep{jameson64} as well as reflection of wave packets \citep{alboussiere11}.

In this paper we elaborate on the remark that reflection of Alfvén waves is controlled not only by the boundary condition, but also by the magnetic Prandtl number of the fluid in which they propagate \citep[see][p. 23,24]{jameson61}. 
In the next section, we discuss the governing equations for one dimensional Alfvén waves and the associated boundary conditions for a solid and electrically insulating wall.
We remark that for $Pm=1$ all the energy of the incident Alfvén wave is dissipated in a boundary layer, resulting in no reflected wave.
In the following section, we change geometry to further emphasize our point and briefly present a direct numerical simulation of propagation and reflection of Alfvén wave in a non rotating spherical shell.
That introduces the section devoted to the geophysical application, where we investigate torsional Alfv\'en waves in the Earth's core, modeled as a rapidly rotating spherical shell, calculating the energy loss on reflection at the Equator as a function of $Pm$.
Finally, we discuss the implications concerning the ability of geodynamo simulations to produce torsional eigenmodes and waves which are expected in the Earth's core.

\section{Reflection of one-dimensional Alfvén waves}

We introduce the problem through the example of Alfvén waves, transverse to a uniform magnetic field in an homogeneous and electrically conducting fluid, hitting a solid wall perpendicular to the imposed magnetic field \citep{roberts67}.
The imposed uniform magnetic field $B_0$ is along the $x$-axis, while the induced magnetic field $b(x,t)$ and the velocity field $u(x,t )$ are transverse to this field, along $y$.
Assuming invariance along $y$ and $z$ axes, the problem reduce to a 1-dimensional problem, $u$ and $b$ depending only on $x$.
Projecting the Navier-Stokes equation and the induction equation on the $y$ direction (on which the pressure gradient and the non-linear terms do not contribute), one obtains the following equations:
\begin{align}
\partial_t u = & \frac{B_0}{\mu_0 \rho} \partial_x b  + \nu \partial_{xx} u \label{eq:momentum} \\
\partial_t b = & B_0 \partial_x u + \frac{1}{\mu_0 \sigma} \partial_{xx} b \label{eq:induction}
\end{align}
where $\mu_0$ is the magnetic permeability, $\rho$ is the fluid density, $\nu$ the kinematic viscosity, and $\sigma$ the electrical conductivity.

\subsection{Elsasser variables}

Introducing the two Elsasser variables $h_\pm = u \pm b/\sqrt{\mu_0\rho}$, the equation of momentum (\ref{eq:momentum}) and the equation of magnetic induction (\ref{eq:induction}) can be combined into 
\begin{equation}
\partial_t h_\pm \, \mp V_A \partial_x h_\pm  - \frac{\eta+\nu}{2} \partial_{xx} h_\pm = \frac{\nu-\eta}{2} \partial_{xx} h_\mp
\end{equation}
where $V_A = B_0/\sqrt{\mu_0\rho}$ is the Alfvén wave speed, and $\eta = (\mu_0 \sigma)^{-1}$ is the magnetic diffusivity.
It is already apparent that when $\nu=\eta$, the right hand side of the previous equation vanishes, in which case $h_+$ and $h_-$ are fully decoupled.
One can also show that $h_-$ travels in the direction of the imposed magnetic field, while $h_+$ travels in the opposite direction.

Introducing a length scale $L$ and the time-scale $L/V_A$, the previous equations take the following non-dimensional form:
\begin{equation}
\partial_t h_\pm \, \mp \partial_x h_\pm  - \frac{1}{S} \partial_{xx} h_\pm = \frac{1}{S}\frac{Pm-1}{Pm+1} \partial_{xx} h_\mp
\label{eq:prop-alf}
\end{equation}
where the Lundquist number $S$ and the magnetic Prandtl number $Pm$ are defined as:
\begin{align*}
S  &= \frac{2 V_A L}{\eta + \nu} & 		Pm &= \frac{\nu}{\eta}
\end{align*}
The propagation of Alfvén wave requires that the dissipation is small enough for the wave to propagate. This is ensured by $S \gg 1$.

The fact that $(Pm-1)/(Pm+1) = -(Pm^{-1}-1)/(Pm^{-1}+1)$ establishes a fundamental symmetry of these equations: when changing $Pm$ into $Pm^{-1}$, only the sign of the coupling term (right hand side of equations \ref{eq:prop-alf}) changes.

\subsection{Physical boundary conditions and reflection of Alfvén waves}
\label{sec:rfl_simple}

These equations must be completed by boundary conditions.
We assume that the wall is electrically insulating, and that the fluid velocity vanishes at the solid boundary (no-slip boundary condition),
which translate to $b=0$ and $u=0$, leading to $h_\pm = 0$.

For $Pm=1$ the equations for $h_+$ and $h_-$ are fully decoupled, regardless of the value of $S$:
\begin{equation}
\partial_t h_\pm = \pm \partial_x h_\pm + \frac{1}{S} \partial_{xx} h_\pm
\end{equation}
In addition, for an insulating solid wall, the boundary condition $h_\pm = 0$ does not couple $h_+$ and $h_-$ either. As a result, reflection is not allowed at an insulating boundary when $Pm=1$, because reflection requires change of traveling direction, and thus transformation of $h_+$ into $h_-$ and vice versa.
The energy carried by the wave has to be dissipated in the boundary layer.

For $Pm \neq 1$ the equations are coupled: for very small diffusivities (that is large Lundquist number $S$), the coupling will be effective only in a thin boundary layer.
In addition the coupling will be more efficient as $Pm$ is further from $1$.
This gives a mechanism for reflection of Alfvén waves on an insulating boundary when $Pm \neq 1$.
Before giving a numerical illustration, it is instructive to consider the boundary conditions in the two limits $Pm=0$ and $Pm=\infty$, with $S \gg 1$ (dissipationless interior).

In the limit $Pm=0$, there is no viscous term and the boundary condition, at the wall $x=x_0$, reduces to
\begin{equation}
b(x_0,t)=0 \quad \Rightarrow \quad h_+(x_0,t)=h_{-}(x_0,t).
\end{equation}
There is perfect reflection. The incident (+) and reflected (-) waves have equal velocities and opposite magnetic fields.
This also corresponds to a stress-free boundary condition for the velocity field in combination with an insulating wall (infinitely small vorticity sheet at the wall), leading to perfect reflection regardless of the value of $Pm$ used in equation \ref{eq:prop-alf}.
In this case the boundary condition for the velocity field is $\partial_x u = 0$, which translates into $\partial_x (h_+ + h_-)=0$ and $h_+-h_-=0$, effectively coupling $h_+$ and $h_-$.

In the limit $Pm=\infty$, the boundary condition , at the wall $x=x_0$, reduces instead to
\begin{equation}
u(x_0,t)=0 \quad \Rightarrow \quad h_+(x_0,t)= -h_{-}(x_0,t).
\end{equation}
The incident and reflected waves have opposite velocities and equal magnetic fields.
This also corresponds to a no-slip boundary condition for the velocity field in combination with a perfectly conducting wall (infinitely small current sheet at the wall), leading to perfect reflection regardless of the value of $Pm$ used in equation \ref{eq:prop-alf}.
In this case the boundary condition for the magnetic field is $\partial_x b = 0$, which couples $h_+$ and $h_-$.

Another combination of boundary conditions inhibits reflection for $Pm=1$: for a stress-free ($\partial_x u = 0$) and perfectly conducting wall ($\partial_x b = 0$), which translates into $\partial_x h_+ = 0$ and $\partial_x h_- = 0$, the fields $h_+$ and $h_-$ are decoupled, as for a no-slip insulating wall.
Note finally that a wall with finite conductivity will allow some weak reflection, as illustrated by figure \ref{fig:taw_space_time}h.

\subsection{Numerical simulations}

\begin{figure}
\centerline{\includegraphics[width=\columnwidth]{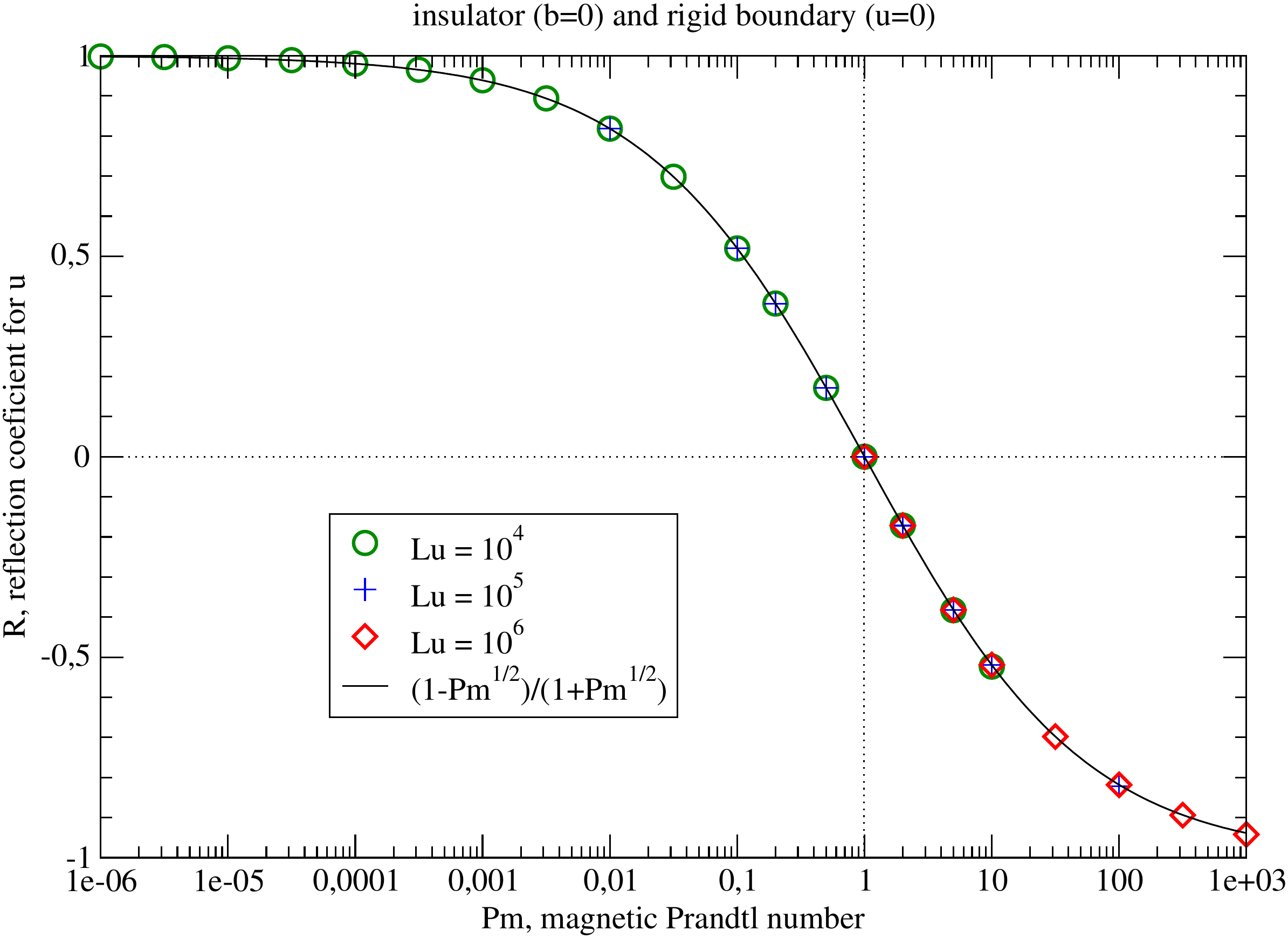}}
\caption{\label{fig:R1D} Reflection coefficient for a one-dimensional Alfvén wave packet hitting an insulating boundary with normal incidence, as a function of $Pm$ and for different magnetic Lundquist numbers $Lu = V_aL/\eta$.
The theoretical value for plane waves $R(Pm) = (1-\sqrt{Pm})/(1+\sqrt{Pm})$ fits the numerical simulation results perfectly.}
\end{figure}

We have performed a numerical simulation in a channel $0\le x\le x_0$ with a one-dimensional finite difference scheme.
The Lundquist number is chosen large enough so that dissipation can be neglected in the interior.
The boundary conditions were set to be electrically insulating and no-slip.
The grid is refined next to the boundaries, in order to have at least 4 points in each boundary layer, which are Hartmann layers of thickness $\delta =  \sqrt{\nu\eta}/V_A$ (see appendix \ref{sec:r_theory}).

From the simulation of the traveling wave, we compute the transmission coefficient as the ratio of the velocity amplitude of the reflected and incident waves for different values of $Pm$ and $S$.
The results are reported on figure \ref{fig:R1D}.

As expected, there is full  dissipation for $Pm=1$ and energy conservation for $Pm \gg 1$ or $Pm \ll 1$.
Furthermore, the reflection coefficient $R$ is independent of $S$, and exhibits the expected symmetry $R(Pm^{-1}) = -R(Pm)$.
The measured values of $R$ match perfectly the theoretical reflection coefficient $R(Pm) = (1-\sqrt{Pm})/(1+\sqrt{Pm})$ derived for plane waves, because $R$  depends neither on the pulsation $\omega$, nor on the wave number $k$ (see appendix \ref{sec:r_theory}).

\section{Reflection of a localized Alfvén wave packet on a spherical boundary}
\label{sec:marie}


\begin{figure*}
\centerline{\includegraphics[width=0.23\textwidth]{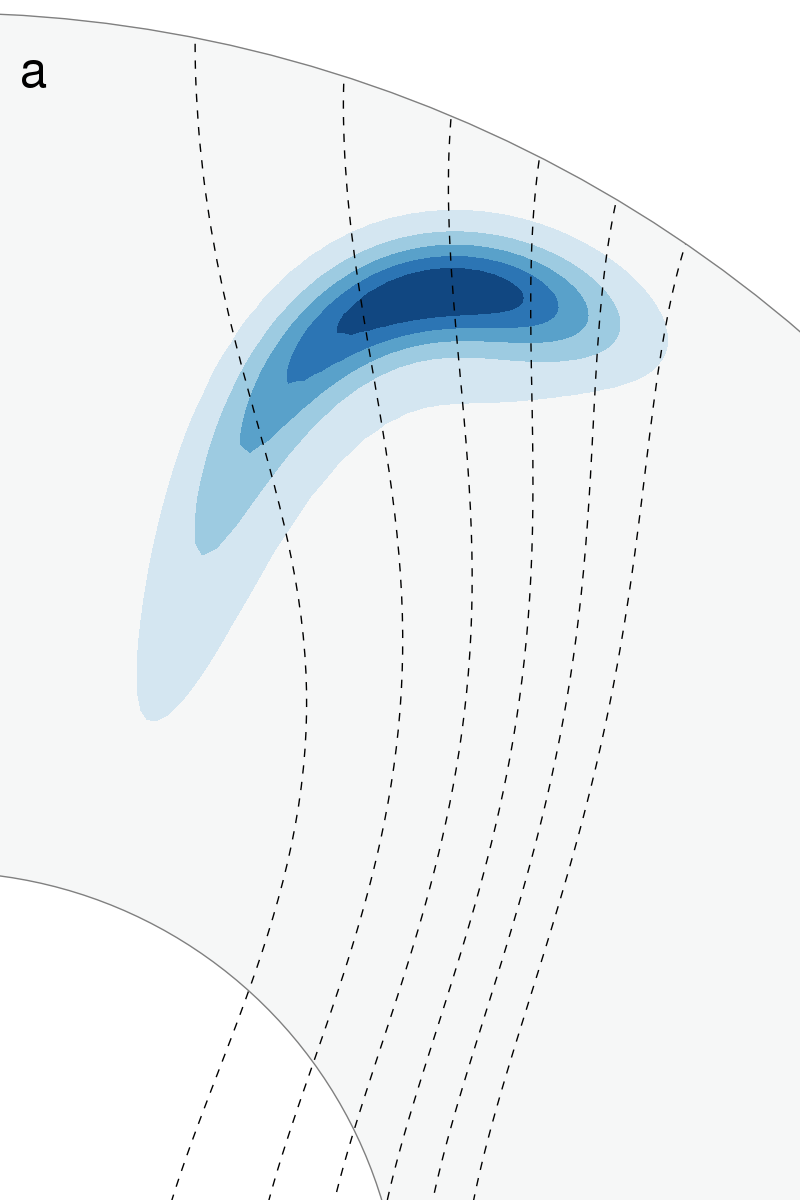}  \includegraphics[width=0.23\textwidth]{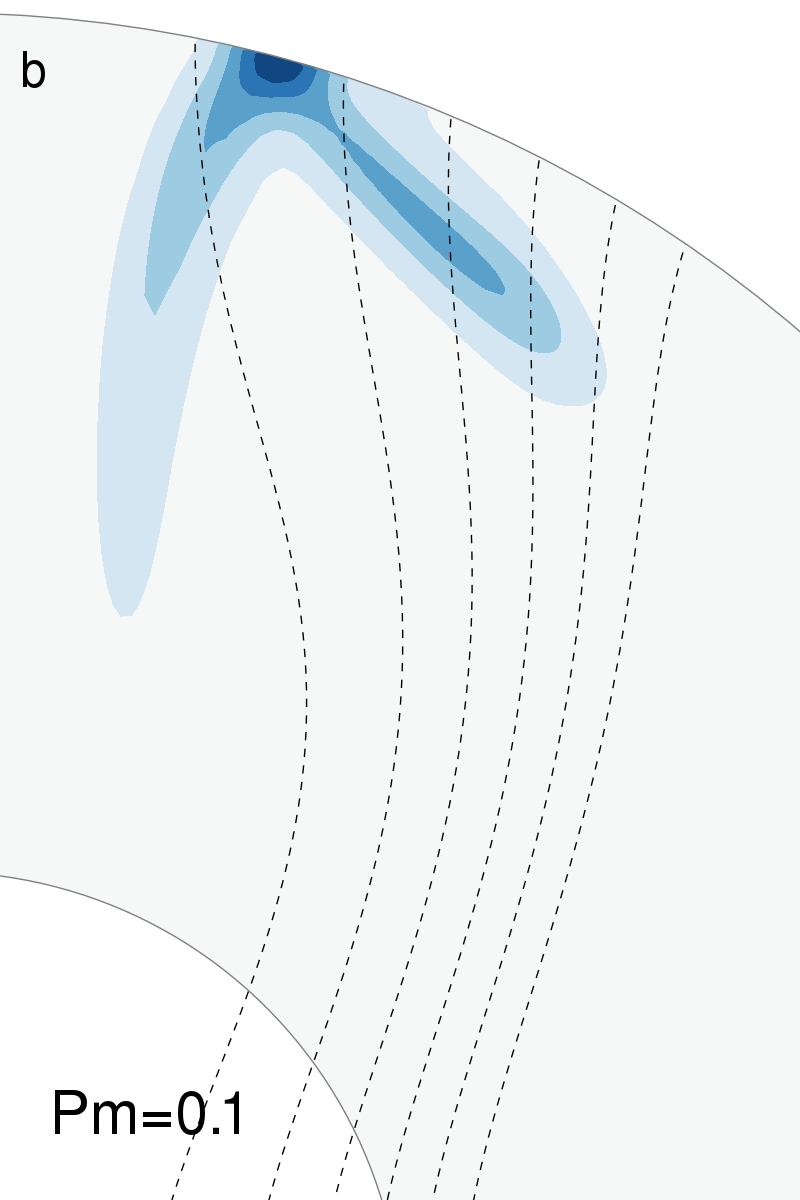}
\includegraphics[width=0.23\textwidth]{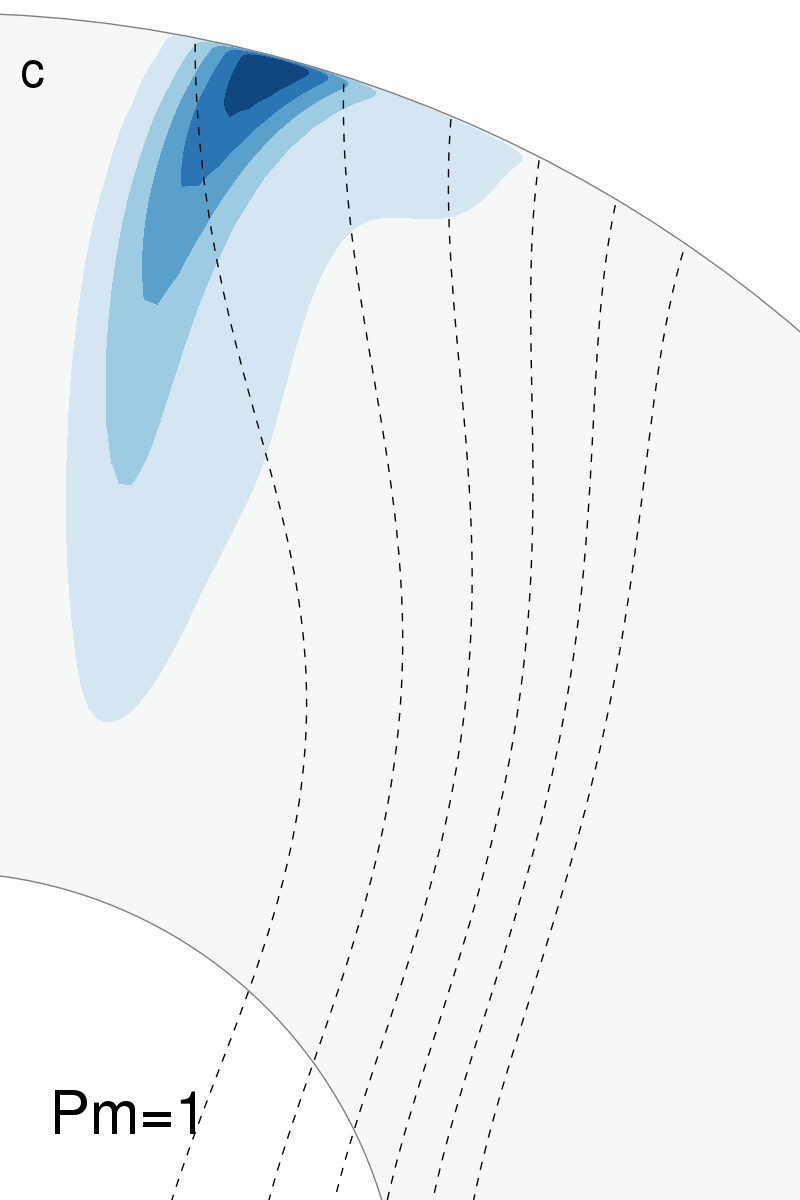} \includegraphics[width=0.23\textwidth]{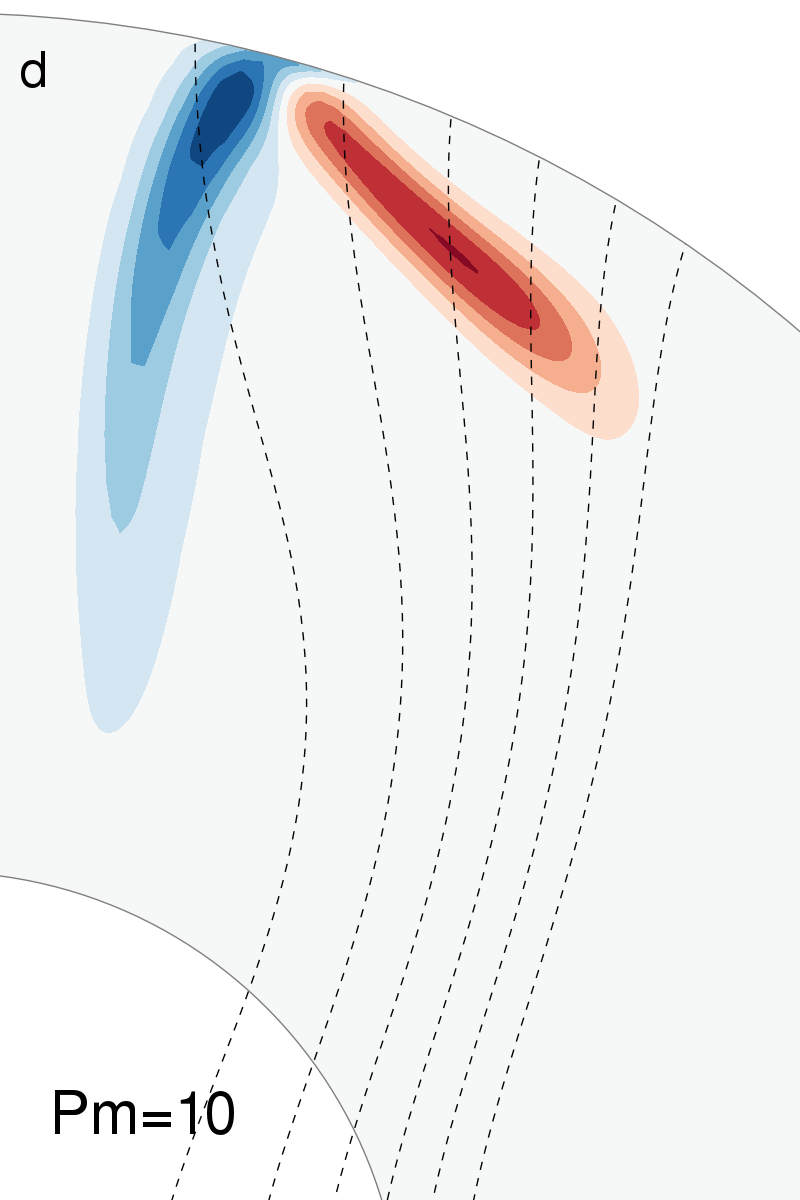} }
\caption{\label{fig:marie} Snapshot of the azimuthal velocity component of Alfvén waves propagating in a non-rotating spherical shell. The dashed-lines are the imposed magnetic field lines.
From left to right: (a) the incoming waves traveling from the inner shell to the outer shell along magnetic field lines;
(b) case $Pm=0.1, S=1800$ showing reflection with the same sign;
(c) case $Pm=1, S=1000$ with total absorption at the wall;
(d) case $Pm=10, S=1800$ showing reflection with opposite sign.}
\end{figure*}

The peculiar case where no reflection occurs is not specific to the planar, one-dimensional ideal experiment.
Here, we run an axisymmetric simulation in a spherical shell permeated by a non-uniform magnetic field, without global rotation.
The imposed magnetic field is the same as in \cite{jault08}, and is represented by the dashed field lines of figure \ref{fig:marie}.
Contrary to the simplest case of the previous section, it is a non-uniform magnetic field, which is not perpendicular to the boundaries.
The observed behaviour of Alfvén wave packets hitting the curved boundaries should therefore apply to many systems.

The numerical pseudo-spectral code is the one used in \cite{gillet11}, but restrained to axisymmetry.
It uses the SHTns library \citep{shtns} for spherical harmonic expansion (Legendre polynomials) in the latitudinal direction, and second order finite differences in radius with many points concentrated near the boundaries.
It time-steps both induction and momentum equation in the sphercial shell using a semi-implicit Crank-Nicholson scheme for the diffusive terms, while the coupling and (negligible) non-linear terms are handled by an Adams-Bashforth scheme (second order in time).
The number of radial grid points is set to 500 and the maximum degree of Legendre polynomials to 120.

The Alfvén wave packets are generated mechanically by spinning the conducting inner core for a very short duration (compared to the Alfvén propagation time).
Since the imposed magnetic field strength is not uniform, the wave front deforms as it propagates along the field lines.
When the wave packet hits the outer insulating spherical shell, it does reflect and propagates back towards the inner shell for $Pm = 0.1$ and $Pm=10$ but there is no reflection for $Pm=1$.
This is illustrated by the snapshots of figure \ref{fig:marie}.

\section{Reflection of Torsional Alfvén waves}

Finding evidence of propagation of Torsional Alfv\'en Waves (TAW) in the Earth's fluid core may open a window on the core interior. 
Properties of TAW in the Earth's core have thus been thoroughly investigated after the initial study of \cite{braginski70}.
They have been recently reviewed by \cite{jault03} and \cite{roberts11}.

\subsection{Model of Torsional Alfvén waves}

In order to model TAW, magnetic diffusion and viscous dissipation are neglected in the interior of the fluid. The Earth's fluid core is modeled as a spherical shell of inner radius $r_i$, outer radius $r_o$ and rotation rate $\Omega$.
Rapid rotation introduces an asymmetry between the velocity and magnetic fields and makes the velocity geostrophic, provided that 
$\lambda\equiv V_A/\Omega r_o\ll 1$ \citep{jault08}.
Note that the Lehnert number $\lambda$ is about $10^{-4}$ in the Earth's core.
Geostrophic velocity in a spherical shell consists of the rotation
$\omega_g(s)$ of nested cylinders centered on the rotation axis. It thus depends only on the distance $s$ from the rotation axis.
A one-dimensional wave equation for the geostrophic velocity $s \omega_g(s)$ is obtained after elimination of the magnetic field 
$b$: 
\begin{equation}
{L}\frac{\partial^2\omega_g(s)}{\partial t^2}=\frac{\partial}{\partial s} \left({L}\tilde{V}_A^2\frac{\partial\omega_g(s)}{\partial s}\right)
\label{eq:ot}
\end{equation}
\noindent with $L=s^3 H(s)$ and $H(s)$ the half-height of the geostrophic cylinders, and $\tilde{V}_A^2$ involves only the $z$-average of the squared $s$-component of the imposed magnetic field.
\cite{braginski70} derived (\ref{eq:ot}) rigorously in the  geophysical case for which the viscous Ekman layer is thin compared to the magnetic diffusion layer located at the top and bottom rims of the geostrophic cylinders.
This condition amounts to $Pm \lambda \ll 1$.
Then, the velocity remains geostrophic in the magnetic diffusion layer.
We have written the equation (\ref{eq:ot}) in its simplest form, when the imposed magnetic field is axisymmetric, the mantle is insulating and Ekman friction at the rims of the geostrophic cylinders is neglected.
The equation (\ref{eq:ot}) needs to be completed by two boundary conditions, which can be derived when either $Pm \ll 1$ or $Pm \gg 1$.

Interestingly, the equation (\ref{eq:ot}) may be valid in the limit $Pm  \ll 1$ but also in the limit $Pm  \gg 1$ (provided $Pm \lambda \ll 1$).
In the specific case $Pm\ll 1$, the appropriate boundary condition on the geostrophic velocity at the equator (on the inner edge of the Hartmann boundary layer) can be inferred from the boundary condition on the magnetic field.
For an insulating outer sphere, it yields $\partial_s \omega_g = 0$ which corresponds to a stress-free boundary, as in the one dimensional wave case with $Pm \to 0$.
In the case $Pm  \gg 1$, the appropriate boundary condition is $\omega_g=0$ as the angular velocity of the outermost geostrophic cylinder is immediately synchronized with the rotation of the solid outer sphere in the course of a spin-up experiment.
This is equivalent to a no-slip boundary, as for the one dimensional wave case with $Pm \to \infty$.

\subsection{Normal modes}
\label{sec:normal}

Assuming that $\omega_g$ varies with time as $e^{i c t}$, the equation (\ref{eq:ot}) can be transformed into a normal mode equation:
\begin{equation}
-c^2 \omega_g(s)=\frac{1}{L}\frac{\partial}{\partial s} \left({L}\tilde{V}_A^2\frac{\partial\omega_g(s)}{\partial s}\right)
\label{eq:nm}
\end{equation}
\noindent
Transmission and reflection of TAW on the geostrophic cylinder tangent to the inner core set a special problem that we do not address here. As an intermediate step, we simply illustrate our discussion with results for the full sphere case, imposing  $\partial_s \omega_g |_{s=\varepsilon} = 0$, with
$\varepsilon\ll 1$ (we have checked the convergence of the numerical results as $\varepsilon\rightarrow 0$).
 It is of interest to write the solution of this equation in the case $c=0$ and $\tilde{V}_A$ uniform:
\begin{equation}
\omega_g(s)=\frac{1}{2}\alpha_1\left( -\frac{\sqrt{1-s^2}}{s^2}-\log\left(\sqrt{1-s^2} + 1\right)+\log(s)\right) +\alpha_2
\end{equation}
\noindent
A non-zero solution (uniform rotation $\omega_g(s)=\alpha_2$) exists for the boundary condition $\partial_s \omega_g |_{s=r_o} =0$ but not for the condition $\omega_g|_{s=r_o} =0$ that applies when $Pm  \gg 1$. We are interested in this latter case, despite its lack of geophysical realism, as contrasting the two boundary conditions sheds light on the nature
of the constraint $\partial_s \omega_g |_{s=r_o} =0$ that has always been used in TAW studies.

In the general case ($c\neq 0$, non-uniform $\tilde{V}_A$), it remains easy to calculate numerically a solution of (\ref{eq:nm}) for $0 < s < r_o$.
We have successfully checked our numerical results against the eigenvalues listed in the table C1 of \cite{roberts11}, that have been obtained analytically for 
$\partial_s \omega_g |_{s=r_o} = 0$ and $\tilde{V}_A=1$.
Then, the first eigenvalues are $(0, 5.28, 8.63, 11.87, 15.07, ..)$, while in the case $\tilde{V}_A=1$ and $\omega_g|_{s=r_o}=0$ they are $(2.94,6.35,9.58,12.78, 15.95, ..)$. In the latter case, we recover our previous observation that $0$ is not an eigenvalue.

In contrast with an often-made statement \citep{buffett98,jault03,roberts11}, the study of the case $Pm \gg 1$  shows that it is not required to have 
\mbox{$\partial_s \omega_g |_{s=r_o} = 0$} to obtain solutions with bounded values of $\omega_g$  for $ s \le r_o$.
On the other hand, the singularity of $\partial_s L$ at $s=r_o$ implies a singularity of $\partial_s \omega_g$ (which is $O((1-s)^{-1/2})$ as $s \rightarrow 1$) . That points to significant viscous dissipation once the viscous term is reintroduced.

When $Pm$ is neither very small nor very large, it is not possible to separate the interior region (where (\ref{eq:nm}) applies) and the Hartmann boundary layer.

We can conclude the discussion of normal modes by noting that the solutions for the two cases $Pm \ll 1$ and $Pm \gg 1$ differ in a significant way at the equator. In both cases, solutions with bounded values of $\omega_g$  in the interval $\left[0,r_o\right]$ and satisfying the appropriate boundary conditions are obtained. However,  reintroducing dissipation  modifies the eigensolutions in the vicinity of the equator and the eigenvalues in the second case only.

\subsection{Numerical experiments}

In order to determine the reflection coefficient of TAW at the equator of the outer shell, we use a set-up that resembles the Earth's core.
The code is the same as the one described in section \ref{sec:marie}, but this time with imposed global rotation.
The total number of radial points is typically 1200 and the maximum degree of Legendre polynomials is set to 360.

For reflection to occur, there must be a non-zero imposed magnetic field $B_s$ at the equator. 
Hence we set the simplest potential quadrupolar field (generated from outside the sphere): $B_s=B_0s$, $B_z=2B_0z$ and $B_\phi = 0$.
This ensures a local traveling speed  $V_A(s) = B_s(s)/\sqrt{\mu_0\rho}$ that is large near the reflection point ($s=1$).
The Lehnert number is small and always set to $\lambda = V_A/(\Omega r_o) = 5\times 10^{-4}$, so that $\lambda Pm$ is also small.

The initial velocity field is along the azimutal direction $\phi$ and depends only on the cylindrical radius $s$: $u_\phi(s) = s\omega_g(s) = u_0 s \exp(- (s-s_0)^2 / \ell^2)$ with $s_0=0.675$. We used two different width $\ell=0.02$ and $\ell=0.063$.
This initial velocity field splits into a torsional Alfvén wave packet propagating inwards that we do not consider here, and another traveling outwards that we carefully follow and we focus on the reflection of this wave packet at the equator of the outer shell ($s=1$).
The Lundquist number $S$ based on the size of the spherical shell ranges from $6\times 10^2$ to $8\times 10^4$ and the Ekman number $E=\nu/\Omega r_o^2$ and magnetic Ekman number $Em=\eta/\Omega r_o^2$ are both always very low and range from $5\times 10^{-10}$ to $5\times 10^{-7}$ over a wide range of magnetic Prandtl number: from $Pm=10^{-3}$ to $Pm=10^2$.

\begin{figure}
\centerline{\includegraphics[width=0.45\textwidth]{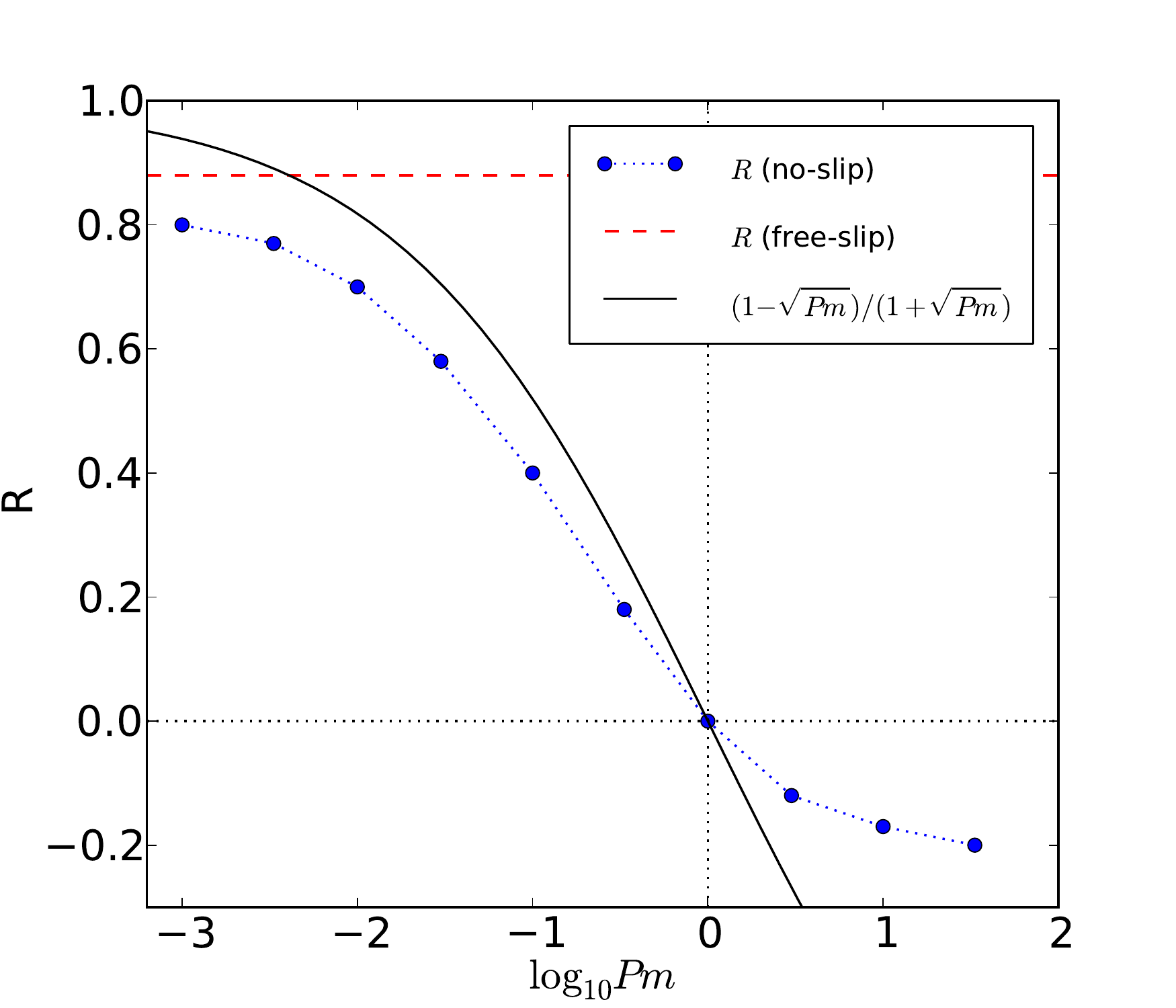}}
\caption{\label{fig:ru_taw}
Reflection coefficient for a torsional Alfvén wave for insulating and no-slip boundary conditions, as a function of $Pm$.
The Lundquist number is always large ($S > 5000$ for $Pm \geq 0.01$ and $S>600$ otherwise).
For reference, the black curve is the planar Afv\'en wave reflection coefficient $(1-\sqrt{Pm})/(\sqrt{Pm}+1)$, and the red line marks the reflection coefficient for a stress-free boundary with $Pm=1$ (corresponding to a no-slip boundary with $Pm \to 0$)}
\end{figure}

\begin{figure*}
\begin{center}
\includegraphics[width=0.45\textwidth]{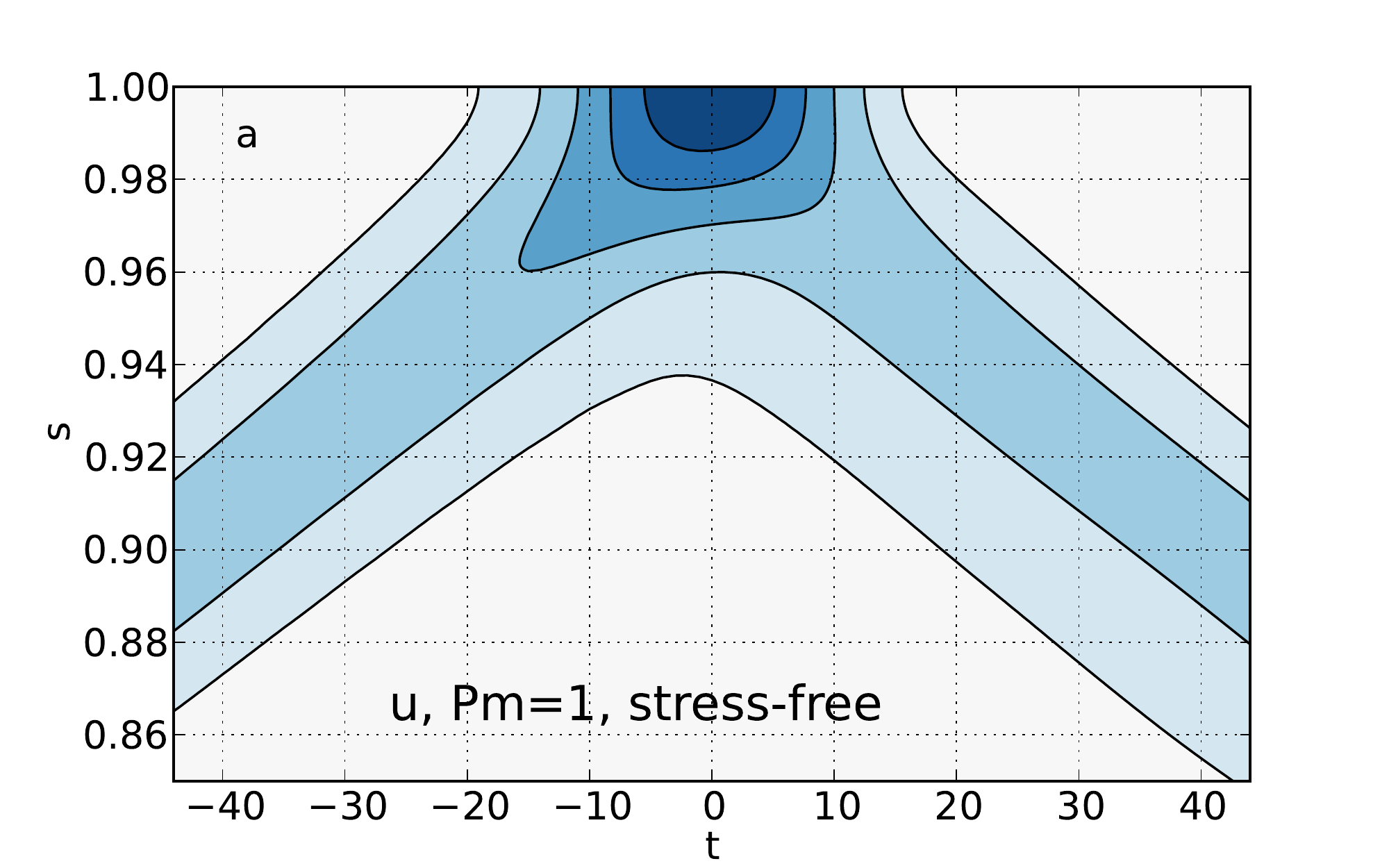}\includegraphics[width=0.45\textwidth]{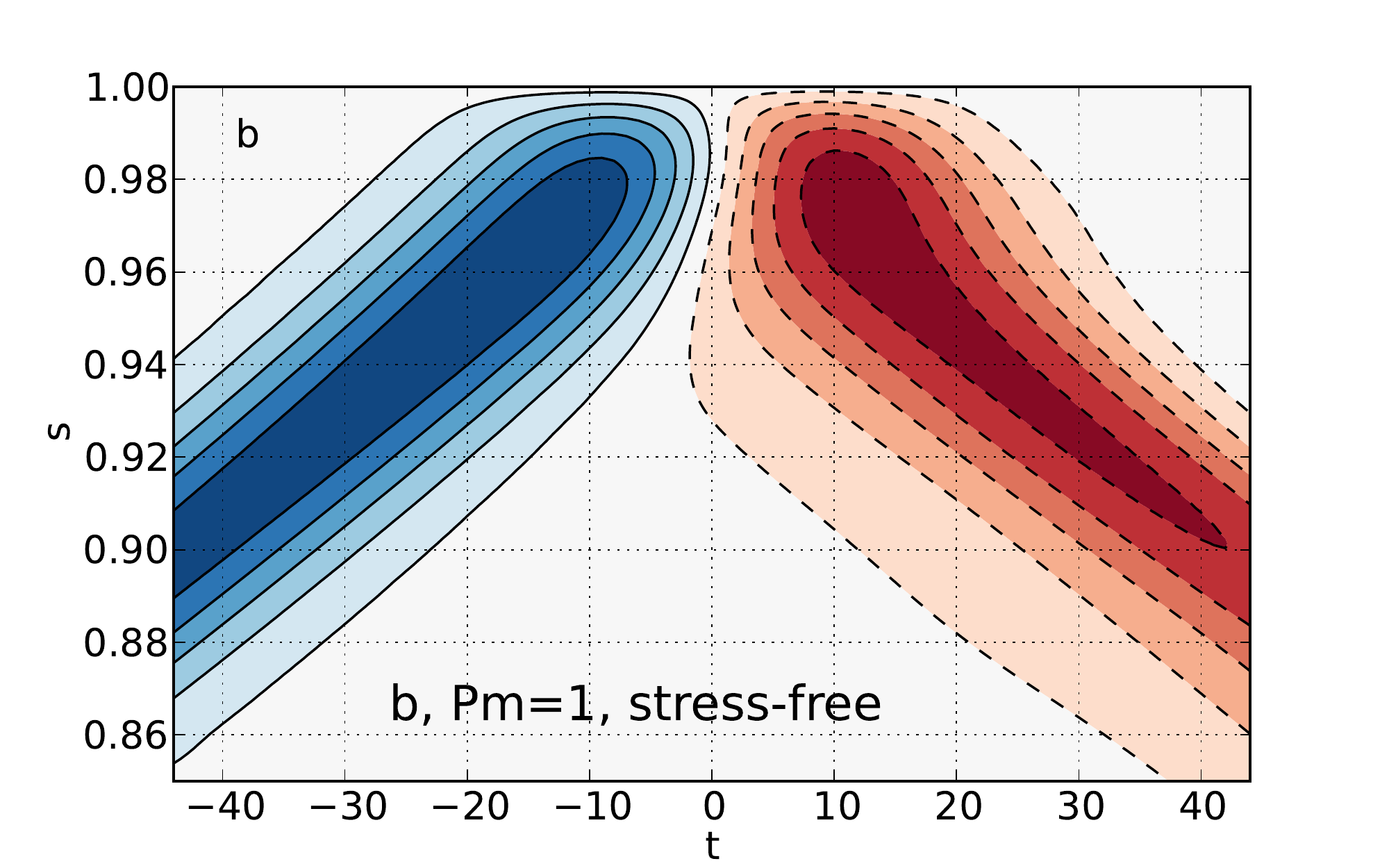}	
\includegraphics[width=0.45\textwidth]{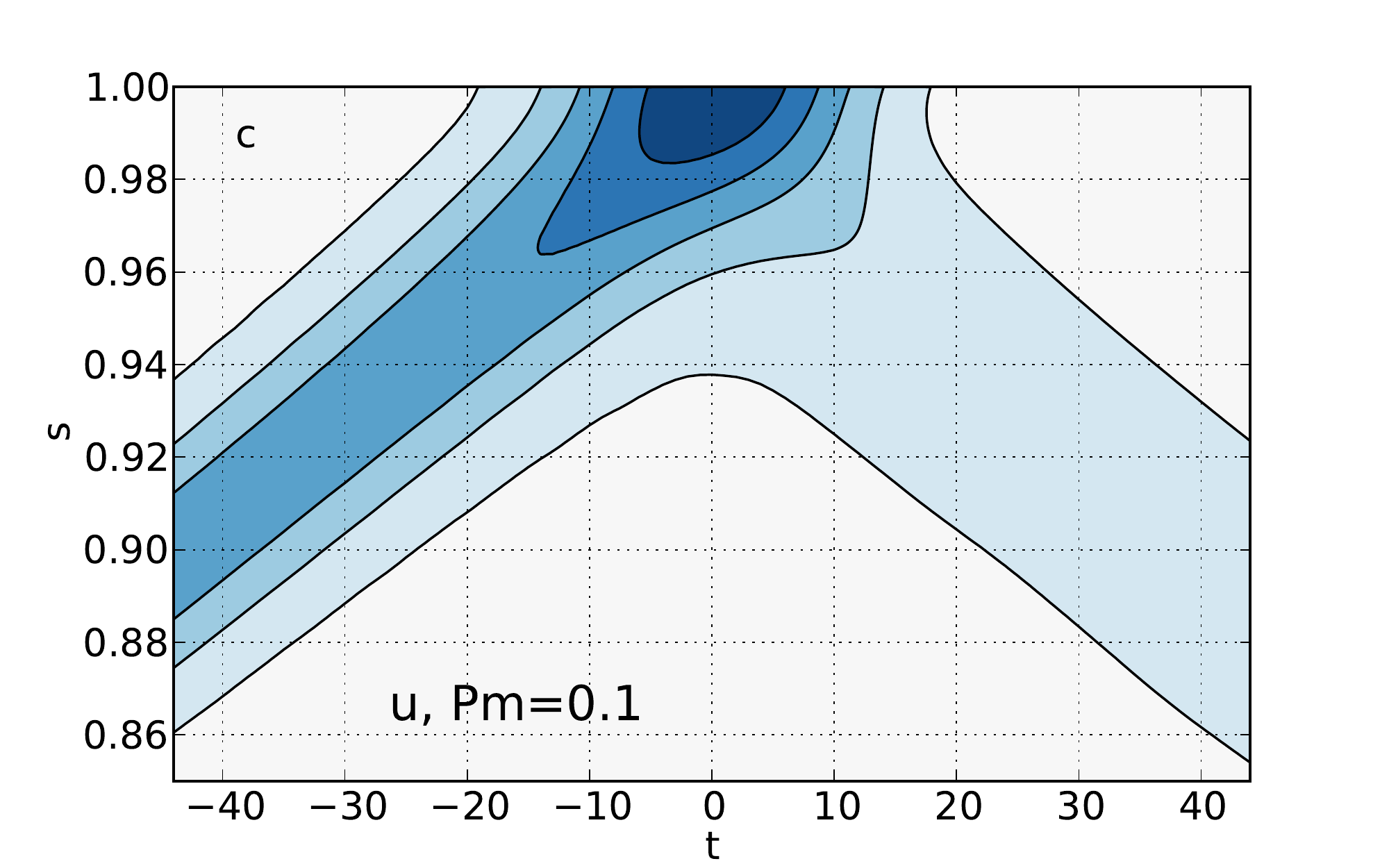}\includegraphics[width=0.45\textwidth]{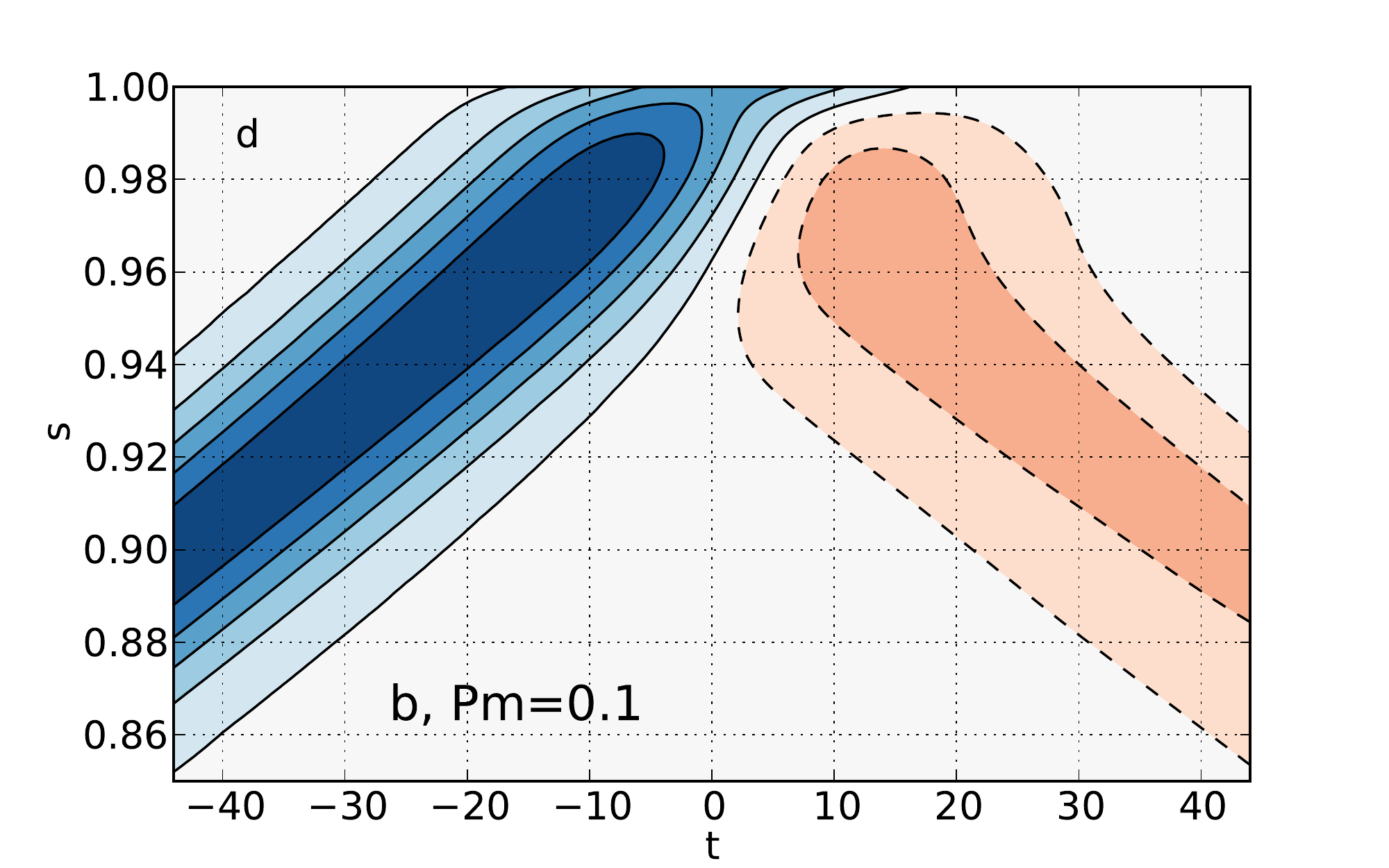}	
\includegraphics[width=0.45\textwidth]{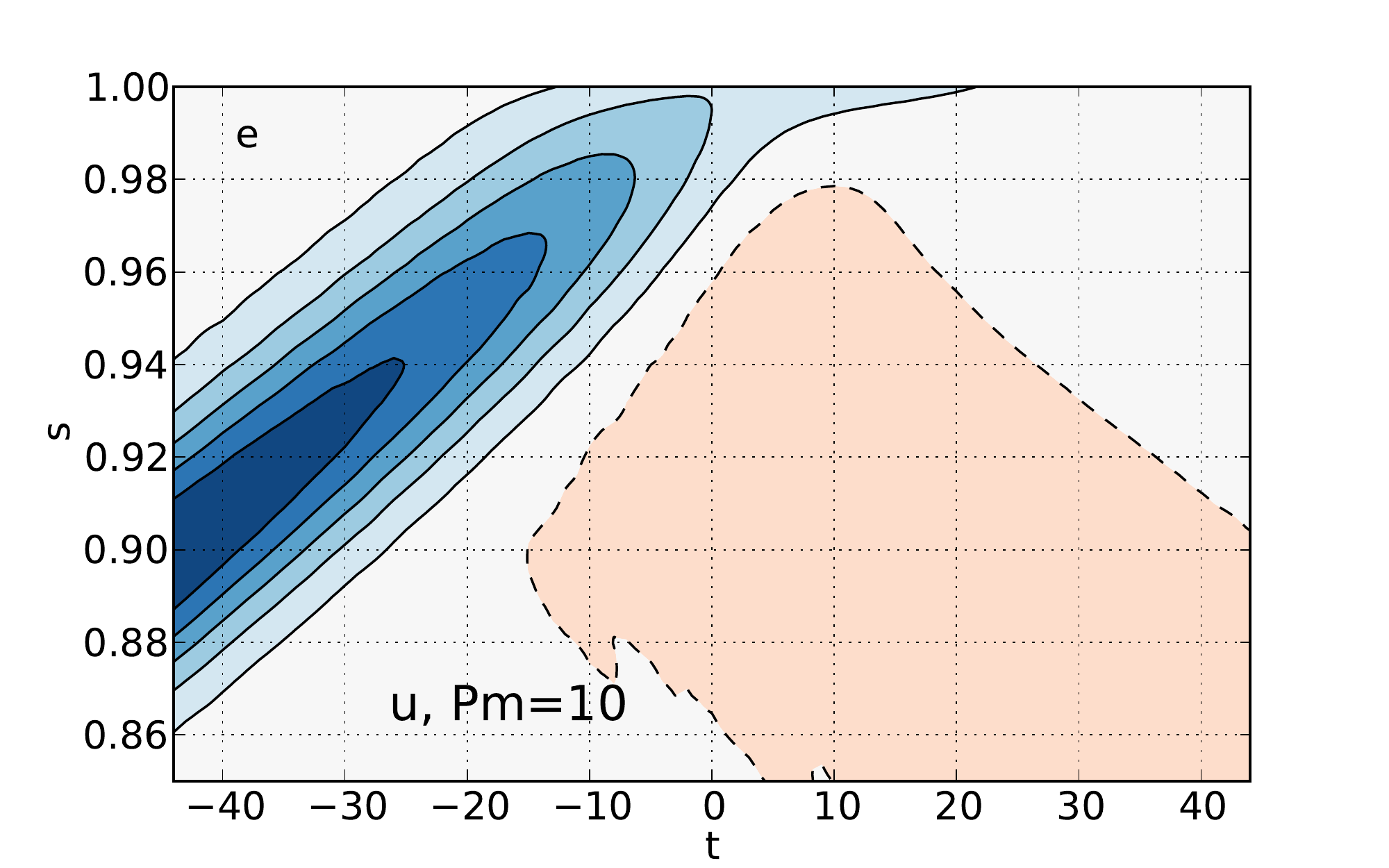}\includegraphics[width=0.45\textwidth]{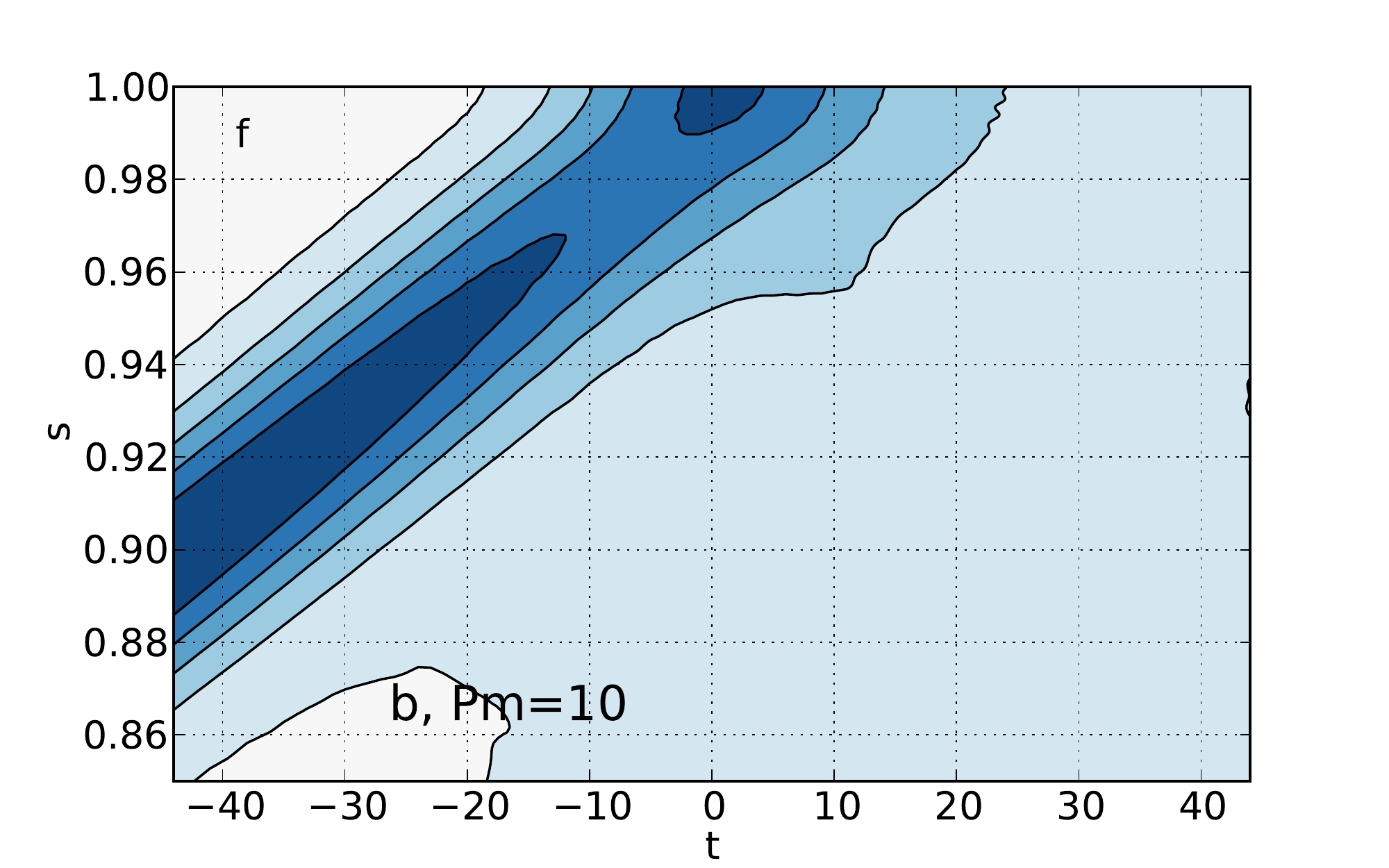}	
\includegraphics[width=0.45\textwidth]{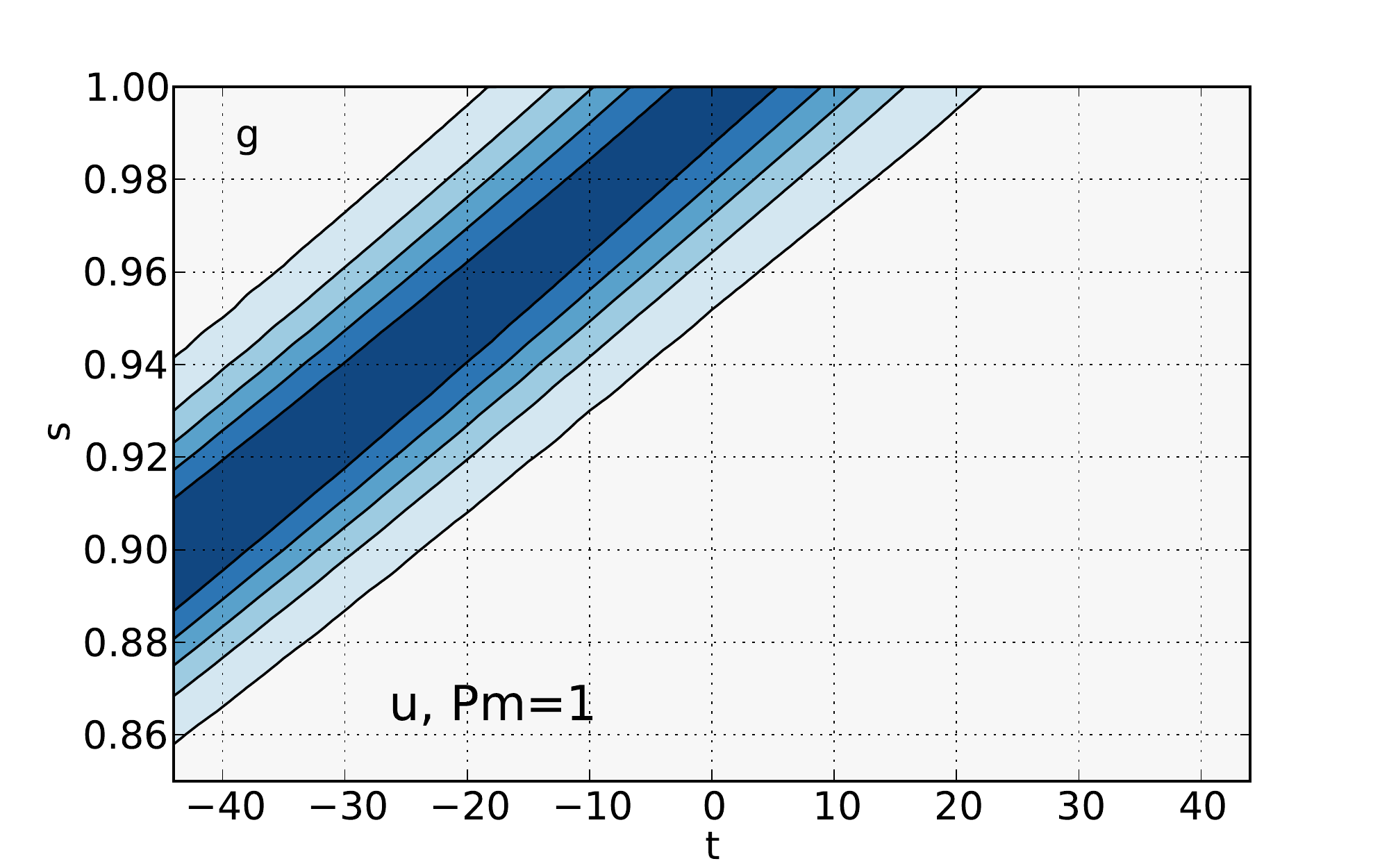}\includegraphics[width=0.45\textwidth]{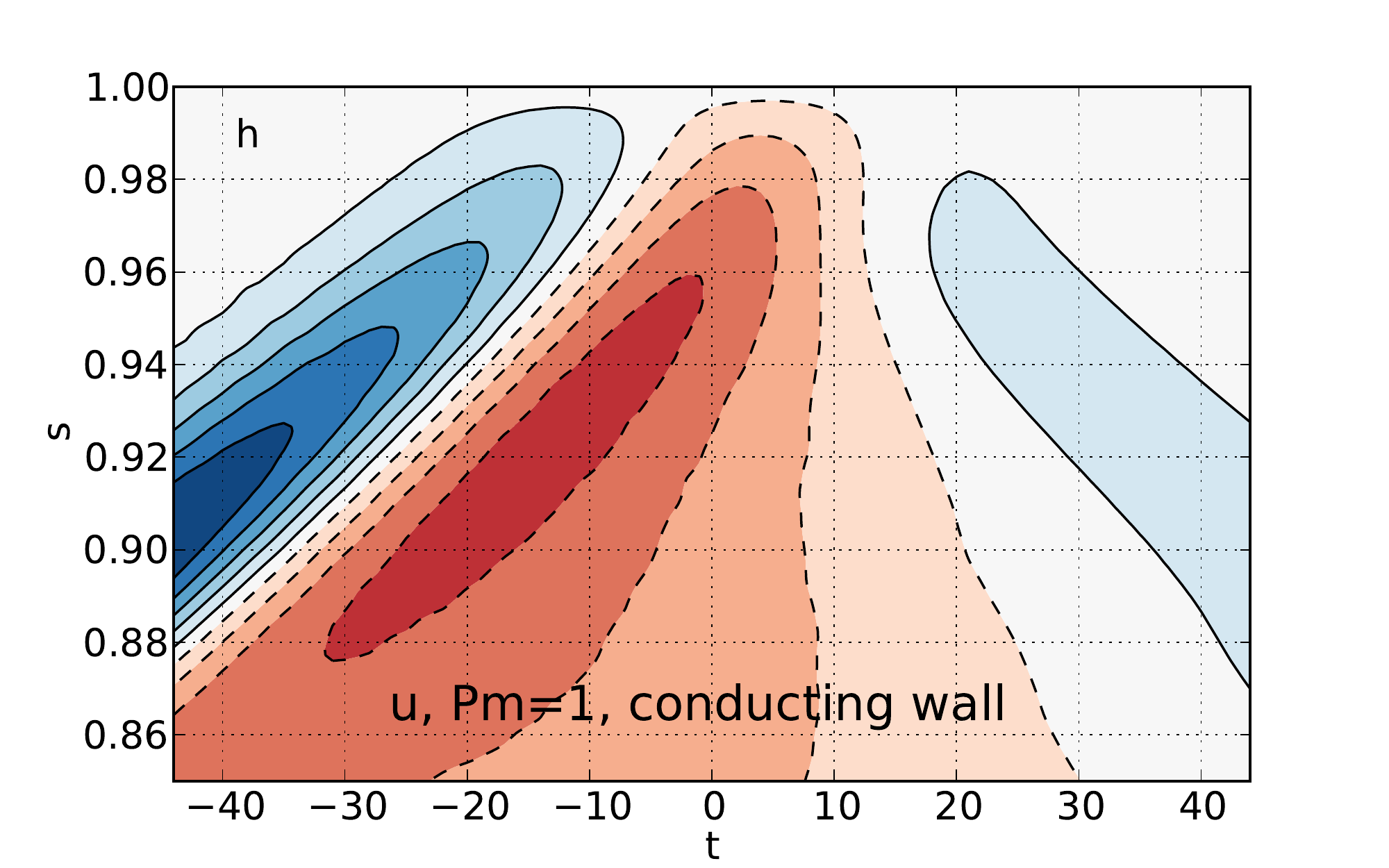}	
\end{center}

\caption{\label{fig:taw_space_time} Space-time diagrams of the reflection of a TAW for $S \simeq 10^4$ and $\ell=0.02$ recorded in the equatorial plane, near the equator.
\emph{Top row}: stress-free boundary with $Pm=1$ ($R=0.88$), (a) the azimuthal angular velocity $u_\phi/s$ and (b) the azimuthal magnetic field $b_\phi$ (changing sign).
\emph{Second row}: No-slip boundary with $Pm=0.1$ ($R=0.40$), (c) the azimuthal angular velocity $u_\phi/s$ and (d) the azimuthal magnetic field $b_\phi$ (changing sign).
\emph{Third row}: No-slip boundary with $Pm=10$ ($R=-0.17$), (e) the azimuthal angular velocity $u_\phi/s$ (changing sign) and (f) the azimuthal magnetic field $b_\phi$.
\emph{Bottom row}: azimuthal angular velocity $u_\phi/s$ for no-slip boundary with $Pm=1$ showing no reflection ($R=0$) for insulating boundary (g), and little reflection when the insulator is replaced by a solid conductive layer (h).
}
\end{figure*}

We measure the extremum of the velocity field in the wave packet before and after the reflection, $a_i$ and $a_r$ respectively, at a fixed radius ($s=0.925$ for $\ell=0.02$ and $s=0.75$ for $\ell=0.063$), from which we compute the corresponding reflection coefficient $R = a_r / a_i$, reported in figure \ref{fig:ru_taw} for an insulating outer shell.
We found no significant dependence with the Lundquist number $S$ or the width of the initial pulse $\ell$ ($R$ varies by less than $0.03$).

As expected from the discussion of Alfvén waves equations, the combination $Pm=1$, no-slip boundary condition and insulating wall corresponds to a special case whereby no reflection at all occurs at the equator (see also fig. \ref{fig:taw_space_time}g).

However, there are differences with the planar case.
First, the reflection coefficient is not symmetric with respect to $Pm=1$, as expected from our discussion of torsional eigenmodes in spherical geometry in the previous section.
For large $Pm$ there is high dissipation and very little reflection compared to low $Pm$.
Second, the reflection coefficient is not as large.

Space-time diagrams of the reflection of the wave at the equator are presented in figure \ref{fig:taw_space_time} for a few representative cases.
The highest reflection coefficient occurs for the stress-free insulating case at $Pm=1$: from $R=0.86$ at $S=1000$ to $R=0.88$ at $S=1.5\times 10^4$.
In this case (fig. \ref{fig:taw_space_time}ab) one can also see the amplification of the velocity field very near the boundary, as the magnetic field must vanish, doing so by producing the reflected wave, just as in the planar case.
This is not a boundary layer, but simply the superposition of the incident and reflected wave (see also appendix \ref{sec:r_theory}).
The Hartmann boundary layer is too small to be seen on these plots, but we checked that its size and relative amplitude for velocity and magnetic fields do match the analytic theory developed in appendix \ref{sec:r_theory}.

For $Pm=0.1$, the reflected wave carries only $16\%$ of the energy, the remaining being dissipated in the boundary layer. The magnetic field changes sign at the reflection, while the velocity keeps the same sign (fig. \ref{fig:taw_space_time}cd).
For $Pm=10$, the reflected energy drops to $3\%$ and the small reflected velocity field has opposite sign, while the magnetic field (barely visible on figure \ref{fig:taw_space_time}) keeps the same sign (fig. \ref{fig:taw_space_time}ef).
During its propagation, the incoming wave is also much more damped than for $Pm=0.1$, even in the case where $S$ or $E$ have comparable values.
This is due to strong dissipation at the top and bottom boundaries, which increases as the wave propagates toward the equator (visible in figure \ref{fig:taw_space_time}e) for $Pm>1$.
This may not be unrelated to the previously discussed singularity for normal modes in the case $Pm>1$.
A consequence of this large dissipation, is the difficulty to clearly identify the reflected wave, and to properly define a reflection coefficient.
The values reported in figure \ref{fig:ru_taw} are thus not very precise for $Pm>1$.

It may also be worth noting that changing the magnetic boundary from insulating to a thin conducting shell allows weak reflection for $Pm=1$ and no-slip velocity (fig. \ref{fig:taw_space_time}h), in agreement with the analysis of the governing equations (section \ref{sec:rfl_simple}).


\subsection{Energy dissipation and normal modes}
\label{sec:nrj}

We want to emphasize that when no reflection occurs, the energy of the wave is dissipated very quickly.
However, for liquid metals ($Pm \ll 1$), only a small amount of the wave energy is absorbed in the event of a reflection, but many successive reflections can lead to significant dissipation.
Using the theoretical reflection coefficient, we can estimate the time-scale of dissipation of an Alfvén wave due to its reflections at the boundaries.
In the case of an Alfvén wave turbulence (many wave packets) in a spherical shell of radius $L$ with homogeneous mean energy $e$, permeated by a magnetic field of rms intensity $B_0$, any wave packet will reach the outer insulating boundary once (on average) in the time interval $L/V_A$.
When it reflects on the boundary, it loses the fraction $1-R^2(Pm)$ of its energy, where $R(Pm)$ is the reflection coefficient (in amplitude).
We can then estimate the dissipation rate of energy $e$ due to this process:
\begin{equation}
	\partial_t e \sim \left( R^2(Pm)-1 \right) \frac{B_0}{L\sqrt{\mu_0 \rho}} \ e
\end{equation}
Hence, the time-scale of dissipation at the boundaries
\begin{equation}
	\tau_s = \frac{L}{V_A} \frac{1}{1-R^2(Pm)}
\label{eq:tau_s}
\end{equation}
which is inversely proportional to the strength of the magnetic field, and depends on the diffusivities only through $Pm$.

We can compare this to the dissipation of Alfvén waves of length scale $\ell$ in the bulk of the fluid: $\tau_v = 2\ell^2/(\eta+\nu)$.
It appears that the length scale $\ell$ where surface and bulk dissipation are comparable is such that
\begin{equation}
	L/\ell = \sqrt{S} \, \sqrt{1-R^2}
\end{equation}
Replacing $R$ by its theoretical expression \ref{eq:R_th} and assuming $Pm \ll 1$, we find for liquid metals
\begin{equation}
	L/\ell = 2 \sqrt{S} \, Pm^{1/4}
\end{equation}
Hence for the Earth's core with $S \sim 10^{4}$, and $Pm \sim 10^{-5}$, the dissipation of Alfvén waves is dominated by the partial absorption at the boundaries for length scales larger than $L/10$.
For numerical simulations of the geodynamo with $S \sim 10^3$ ans $R \sim 0.2$, we have $L/\ell \sim 30$.

These time-scales are also relevant for torsional normal modes. In one dimension, normal modes are a superposition of waves propagating in opposite directions.
Hence, if the dissipation of waves is dominated by their reflection, so will it be for the normal modes.
From the previous estimation of $L/\ell$ in the Earth's core, we expect the dissipation of large wave-length torsional Alfvén waves (the ones that can be observed) to be dominated by the effect of reflection.
Furthermore, in order to detect a normal mode, its dissipation time must be much larger than its period $T=2\pi L(cV_A)^{-1}$.
The pulsation $c$ of the first torsional normal modes are given in section \ref{sec:normal} in Alfvén frequency units, and their dissipation time can be estimated by $\tau_s$ for the large-scale normal modes. We define a quality factor for torsional normal modes by
\begin{equation}
	Q = \frac{\tau_s}{T} = \frac{c}{2\pi} \frac{1}{1-R^2}.
\end{equation}
Presence of normal modes requires $Q \gg 1$.
In the Earth's core, we find $Q_E \simeq 13 \, c$ and for numerical simulations of the geodynamo we find $Q_{sim} \simeq 0.16 \, c$.
Considering the largest modes (with $c \simeq 5$), the torsional oscillations should therefore be present in the Earth's core, but are completely absent even from the best current geodynamo simulations.

\section{Discussion: implication for numerical geodynamo models and the Earth-core}

We showed that numerical simulations conducted for $Pm \sim 1$ cannot adequately reproduce the boundary conditions for torsional Alfv\'en waves in the Earth's core (where $Pm\ll 1$).
The small reflection coefficient observed for TAW (figure \ref{fig:ru_taw}) means that it is hard to observe TAW reflection at the equator in numerical simulations of the geodynamo which currently operate with $0.1 < Pm < 10$ \citep[e.g.][]{takahashi08b, sakuraba09}, where the waves are moreover mixed with thermal convection.

As for possible torsional eigenmodes, it is almost impossible to observe them with such low reflection coefficients.
Unfortunately, that severely limits the ability of geodynamo simulations to exhibit torsional oscillation normal modes, because normal modes require a large reflection coefficient to be observable: their period (of order $L/V_A$) must be much larger than the energy dissipation time $\tau_s$ (see expression \ref{eq:tau_s}).
A few studies have tried to pin down torsional eigenmodes \citep{dumberry03, sakuraba08, wicht10} but even though they report waves propagating with the appropriate speed, they report neither reflection of these waves, nor eigenmodes.

Another issue for geodynamo models with very low diffusivities, is that the part of the energy carried by Alfvén waves (regular or torsional) is dissipated very quickly (on an Alfvén wave time-scale), so that an Alfvén wave turbulence would be damped much faster, and the turbulent state may be far from what we would expect in the Earth's core.

Changing the boundary condition to stress-free simulates the case $Pm = 0$ with a high reflection coefficient ($R=0.88$), but still lower than the planar case.
Even though this may still be problematic to observe eigenmodes, numerical models that use stress-free boundaries \citep[e.g.][]{kuang99, dumberry03, busse06, Sreenivasan2011} are intrinsically much more suited for the study of torsional normal modes.
Quasi-geostrophic dynamo models that can compute dynamo models at very low magnetic Prandtl numbers \citep[$Pm < 10^{-2}$ in][]{Schaeffer06}, could also provide an interesting tool to study torsional oscillations.

In the case of the Earth's core, a recent study \citep{Gillet10} found no clear evidence for reflection at the equator, although this has yet to be confirmed.
One might want to invoke turbulent viscosity (see the contrasted views of \cite{deleplace06} and \cite{buffett07} in a different context) to explain this fact, leading to an effective $Pm$ close to $1$ and inhibiting reflection of torsional Alfvén waves.
This would make numerical models more relevant, but is rather speculative.
A solid conductive layer at the top of the core can also have a damping effect on the propagation and reflection of torsional waves, and we plan to investigate these matters in a forthcoming study.

\begin{acknowledgments}
The numerical simulations were run at the Service Commun de Calcul Intensif de l'Observatoire de Grenoble (SCCI).
We want to thank Mathieu Dumberry and an anonymous reviewer for their help in improving this paper, and Henri-Claude Nataf for useful comments.
\end{acknowledgments}

\appendix
\section{Analytic Alfvén wave solutions in one dimension}
\label{sec:r_theory}

\subsection{Plane wave solutions}

Following \cite[p. 15-18]{jameson61}, we look for plane wave solutions of equations \ref{eq:momentum} and \ref{eq:induction},
substituting $u=U e^{i(\omega t + k x)}$ and $b= \sqrt{\mu_0 \rho}\, B e^{i(\omega t + k x)}$:
\begin{align}
\left(i\omega + \nu k^2 \right) U &= V_A ik B \label{eq:plane_u}\\
\left(i\omega + \eta k^2 \right) B &= V_A ik U \label{eq:plane_b}
\end{align}
which we can combine into
\begin{equation}
\nu\eta \, k^4 + \left( V_A^2 + i\omega(\eta+\nu) \right) \, k^2 - \omega^2 = 0
\end{equation}
for which the exact solutions are:
\begin{equation}
k^2 = -\frac{V_A^2}{2\nu\eta} ( 1 + 2i\epsilon) \left( 1 \pm \sqrt{1 + \frac{4\omega^2 \nu\eta}{V_A^4 (1 + 2i\epsilon)^2}} \right)
\end{equation}
where $\epsilon$ is the reciprocal Lundquist number based on the frequency:
\begin{equation}
\epsilon = \frac{\omega(\eta+\nu)}{2V_a^2}
\end{equation}

In the regime where Alfvén waves do propagate, we have $\epsilon \ll 1$ and also $\omega\sqrt{\nu\eta}/V_A^2 \ll 1$ so we can approximate the square root by its first order Taylor expansion, which leads to two solutions $k_1^2$ and $k_2^2$:
\begin{align}
k_1^2 &= \frac{\omega^2}{V_A^2} ( 1 + 2i\epsilon)^{-1} &
k_2^2 &= -\frac{V_A^2}{\nu\eta} ( 1 + 2i\epsilon)
\end{align}

The solutions $k = \pm k_1 = \pm \omega/V_A (1-i\epsilon)$, correspond to the propagation in both directions of an Alfvén wave at the speed $V_A$ and with attenuation on a length scale $V_A/(\epsilon \omega)$.
The solutions $k = \pm k_2 \simeq \pm i/\delta$ correspond to a Hartmann boundary layer of thickness $\delta \equiv \sqrt{\nu\eta}/V_A$.

Finally, from equation \ref{eq:plane_u} and \ref{eq:plane_b} we know that $U$ and $B$ are related for each $k$ by:
\begin{equation}
\frac{B}{U} = \frac{ikV_a}{i \omega + \eta k^2} = \frac{i \omega + \nu k^2}{ikV_a} \equiv \alpha_k
\end{equation}
and for the solutions $k=\pm k_1$ and $k=\pm k_2$, it reduces to
\begin{align}
\alpha_{\pm k_1} &\simeq \pm 1    &
\alpha_{\pm k_2} &\simeq \pm \sqrt{\frac{\nu}{\eta}} = \pm \sqrt{Pm}
\label{eq:alpha_k}
\end{align}
This means that for the travelling wave solution, $U$ and $B$ have always the same amplitude and the same phase when propagating in the direction opposite to the imposed magnetic field, or opposite phase when propagating in the same direction.
For the boundary layers, in the limit $Pm \ll 1$ they involve the velocity field alone, whereas for $Pm \gg 1$ they involve only the magnetic field.

\subsection{Reflection coefficient at an insulating wall}

In order to derive the reflection coefficient, we consider an insulating wall at $x=0$ with an incoming Afvén wave from the $x>0$ region ($k=+k_1$), giving rise to a reflected wave ($k=-k_1$).
The boundary conditions are matched by a boundary layer ($k=+k_2$) localized near $x=0$
(the solution $k=-k_2$ is growing exponentially for $x>0$ and has to be rejected for this problem).
The solution to this problem reads
\begin{align}
u &= e^{i \omega t}\left[ e^{ik_1 x} + R e^{-ik_1 x} + \beta e^{ik_2x} \right]	\label{eq:sol_u} \\
b &= e^{i \omega t}\left[ \alpha_{k_1} \left( e^{ik_1 x} - R e^{-ik_1 x} \right) + \alpha_{k_2} \beta e^{ik_2x} \right] \sqrt{\mu_0 \rho} \label{eq:sol_b}
\end{align}
where we have taken into account the fact that $\alpha_{-k_1} = -\alpha_{k_1}$ (see eq. \ref{eq:alpha_k}).

The boundary conditions $u=0$ and $b=0$ at $x=0$ lead to:
\begin{align*}
1 + R + \beta &= 0     & \alpha_{k_1}(1-R) + \alpha_{k_2} \beta &= 0
\end{align*}
from which we find the amplitude $\beta$ of the velocity boundary layer contribution, and the reflection coefficient $R$ of the amplitude of the velocity component:
\begin{align*}
\beta &= \frac{-2}{1+\alpha_{k_2}/\alpha_{k_1}} &
R &= \frac{1 - \alpha_{k_2}/\alpha_{k_1}}{1 + \alpha_{k_2}/\alpha_{k_1}}
\end{align*}

We are left to evaluate $\alpha_{k_2}/\alpha_{k_1}$ using equations \ref{eq:alpha_k}, which gives $\alpha_{k_2}/\alpha_{k_1} = \sqrt{\nu/\eta}$ at leading order in $\epsilon$, and thus
\begin{equation}
R = \frac{1 - \sqrt{Pm}}{1 + \sqrt{Pm}}
\label{eq:R_th}
\end{equation}
which is independent of $\omega$ and $V_A$.
In the case $Pm=1$, we then have $R=0$ and $\beta=-1$ which means that no reflection occurs and that the amplitude of the incoming wave is canceled by the boundary layer alone.

It may be worth emphasizing that, although the boundary layer has the same thickness $\delta$ in the velocity and the magnetic field components,
in the limit $Pm \to 0$, we have $\beta \to -2$ and $\alpha_{k_2}\beta \to 0$, so that the boundary layer is apparent only in the velocity field component (eq. \ref{eq:sol_u}),
whereas in the limit $Pm \to \infty$, we have $\beta \to 0$ and $\alpha_{k_2}\beta \to -2$, so that the boundary layer is apparent only in the magnetic field component (eq. \ref{eq:sol_b}).

Finally, we remark that if one sets $\nu=0$ or $\eta=0$ from the beginning in equations \ref{eq:plane_u} and \ref{eq:plane_b}, the solution corresponding to the boundary layer does not exist anymore.

\end{document}